# Chromospheric Flashes in a Solar Pore: Insights from Multi-line Spectropolarimetric Diagnostics

Sandeep Dubey,[1, 2] Christian Beck,[3] Rahul Yadav,[4] Tobias Felipe,[5, 6] and Shibu K Mathew[1]

[1]*Udaipur Solar Observatory, Physical Research Laboratory, Udaipur, Rajasthan, 313001, India*

[2]*Indian Institute of Technology, Gandhinagar, Palaj, Gujarat, 382055, India*

[3]*National Solar Observatory (NSO), 3665 Discovery Drive, Boulder, CO 80303, USA*

[4]*Laboratory for Atmospheric and Space Physics, University of Colorado, Boulder, CO 80303, USA*

[5]*Instituto de Astrofísica de Canarias (IAC), Vía Láctea s/n, E-38205 La Laguna, Tenerife, Spain*

[6]*Departamento de Astrofísica, Universidad de La Laguna, E-38206 La Laguna, Tenerife, Spain*

## ABSTRACT

Solar pores are strongly magnetized regions without a photospheric penumbra, with predominantly vertical magnetic fields. We present a multi-line investigation of flashes in a pore using high-resolution SST observations in Fe I 6302 Å, Ca II (8542 & K), and H-$\beta$, complemented with (E)UV observations from IRIS and AIA. Complementary to bisector analysis, spectral inversions with SIR and NICOLE provided temperature, LOS velocity, and magnetic field stratifications. Flashes, confined to the left half of the pore, exhibited cooler temperatures ($\Delta T \approx 400$ K), stronger magnetic fields ($\Delta B \approx 250$ G), greater inclination ($\sim 25°$ vs $\sim 18°$), and persistent upflows ($\sim 0.5$ km s$^{-1}$) relative to the quiescent pore in the photosphere. Flashes were co-spatial with enhanced 3- and 5-minute power in the photosphere, with only 3-minute power persisting in the chromosphere. Flashes were seen upto 50% line depth in Ca II 8542 Å intensity, but not below, and showed central upflows ($\sim 1$ km s$^{-1}$, $1''-2''$) flanked by strong downflows ($\sim 8$ km s$^{-1}$) in the chromosphere. Associated temperature enhancements reached $\sim 500$ K at log $\tau \approx -5$ and $\sim 2500$ K at log $\tau \approx -6$. Flash spectra displayed a bi-modal velocity distribution, with $\sim 52\%$ showing downflows at log $\tau \approx -5$. Flashes corresponded one-to-one with radially outward running waves (5–15 km s$^{-1}$, amplitude $\sim 1$ km s$^{-1}$) near the pore boundary. Spectral diagnostics revealed strong Ca II (8542 & K) core emission, occasional Stokes V reversals, and broadband H-$\beta$ enhancements. The results suggest that pore flashes are confined to the lower and mid-chromosphere, with little influence on the transition region or corona.

## 1. INTRODUCTION

Pores are strongly magnetized regions on the solar surface that, unlike sunspots, lack a penumbra due to their relatively simple and predominantly vertical magnetic field configuration (Simon & Weiss 1970). Typically ranging in diameter from 1 to 6 Mm, pores can persist for a few hours to more than a day (Tlatov 2023). High-resolution imaging has revealed that, despite their simpler appearance, pores host a variety of fine-scale photospheric structures such as umbral dots (UD) and light bridges (LB) (Hirzberger 2003; Ortiz et al. 2010; Sobotka et al. 2012). Their magnetic field strengths typically range between 600 and 1700 G (Keppens & Martinez Pillet 1996; Suetterlin 1998), with near-vertical inclinations close to the pore centre that increase toward the pore boundary, reaching values of 40°–80° from the vertical. Surrounding the pore in the chromosphere is a superpenumbral structure composed of radially oriented fibrils $5''-20''$ in length and approximately $0.5''$ in width (Sobotka et al. 2013).

Owing to their strong and organized magnetic fields, pores are efficient conduits for the propagation of magneto-hydrodynamic (MHD) waves to higher layers of the atmosphere (Centeno et al. 2009; Stangalini et al. 2011; Cho et al. 2015; Grant et al. 2022). These waves often manifest in the chromosphere as intensity and velocity oscillations, with the umbral regions of sunspots and pores exhibiting a dominant 3-minute oscillation. This oscillatory behavior is thought to originate from the leakage and transformation of global p-mode oscillations (Braun et al. 1987, 1988) and the propagation from the photosphere into the chromosphere of waves whose frequency is above the acoustic cut-off frequency (Centeno et al. 2006). As MHD waves propagate upward through the decreasing density of the atmosphere, they steepen into shocks, producing sudden localized brightenings known as flashes (Beckers & Tallant 1969; Bard & Carlsson 2010; Khomenko & Collados 2015; French et al. 2023). Flashes are highly structured events, with fine-scale sub-arcsecond bright and dark components, and are associated with temperature enhancements of ∼1000 K compared to the surrounding umbra (de la Cruz Rodríguez et al. 2013). While traditionally interpreted as the compression phase of upward-propagating slow magnetoacoustic waves, recent observations have revealed flashes with significant downflows in the chromosphere (Henriques et al. 2017; Bose et al. 2019; Dubey et al. 2024; Felipe et al. 2025). These downflowing flashes, sometimes accompanied by velocity–temperature phase lags indicative of partial wave reflection, suggest a possible role of a chromospheric resonant cavity above the umbra (Felipe et al. 2021). The fraction of such events varies across studies, with standing-wave-like behavior observed in a minority of cases.

Despite extensive studies on sunspots, the occurrence, structure, and dynamics of flashes in pores remain less explored. Given the simpler geometry of pores and their reduced spatial sizes, they offer an opportunity to examine whether similar wave–shock interactions and resonant phenomena occur without the complexity introduced by a penumbra. In this study, we analyzed high-resolution multi-line observations to investigate the properties of flashes in a pore across different layers spanning from the photosphere to the transition region and the solar corona. Sections 2 and 3 outline the observations and methodology used in the analysis, followed by results in Section 4. A comparative discussion of the results in Section 5 is followed by the conclusions of the study in Section 6.

## 2. OBSERVATION

We analyzed line scan observations of a pore in the active region NOAA 12858, observed on August 19, 2021, between 11:57 UT and 13:14 UT. The pore (centred at x≈145″, y≈95″) was located at an approximate heliocentric angle of 15° from the solar disc center. The observed field-of-view (FOV) contained two pores (see Figure 1); one large (≈ 8″ in diameter) and one small (≈ 2″). The larger pore hosted a persistent light bridge throughout the 78-minute observation window.

The line scan observations were acquired in multiple spectral lines using the CRisp Imaging SpectroPolarimeter (CRISP; Scharmer 2006; Scharmer et al.

2008) and the CHROMospheric Imaging Spectrometer (CHROMIS; Scharmer 2017) of the 1-m Swedish Solar Telescope (SST; Scharmer et al. 2003). The Fe I 6302 Å and Ca II 8542 Å line scans were recorded from 11:56 UT for 78 minutes at a cadence of ≈37 s. Additional scans in Ca II K and H-β were available only for the first 8 minutes of the Ca II 8542 Å sequence. The SST data were reduced with the SSTRED pipeline (Löfdahl et al. 2021), which employs multi-object multi-frame blind deconvolution (MOMFBD; Löfdahl 2002; Van Noort et al. 2005) to produce science-ready CRISP and CHROMIS datasets (de la Cruz Rodríguez et al. 2015).

Complementary observations were obtained from the Interface Region Imaging Spectrograph (IRIS; De Pontieu et al. 2014), which provided Slit-Jaw Imager (SJI) intensity images in the 2832 Å and Si IV 1400 Å passband. The IRIS observations started at 10:11:37 UT and ended at 12:15:38 UT, with an overlapping time window of 8 minutes with the SST observations. The IRIS slit was on the left side of both pores observed by the SST. We also used Extreme Ultraviolet (EUV) intensity images from the Atmospheric Imaging Assembly (AIA; Lemen et al. 2012) and the photospheric continuum intensity images from the Helioseismic and Magnetic Imager (HMI; Scherrer et al. 2012) onboard the Solar Dynamics Observatory (SDO; Pesnell et al. 2012). Details of the datasets from all instruments in different lines and filters used in this study are summarised in Table 1.

## 3. METHODOLOGY

The line scan observations in different spectral lines from the SST covered slightly different FOVs, requiring alignment before further analysis. This was achieved by identifying multiple common small-scale features across the FOV. The Ca II K and H-β datasets had a different magnification factor compared to the Fe I 6302 Å and Ca II 8542 Å intensity images. We estimated the magnification factor, applied bi-variate spline interpolation to magnify the Ca II K and H-β images, and aligned them to the Ca II 8542 Å maps using a cross-correlation method (Gratadour et al. 2005).

The IRIS observations had a lower spatial sampling than those from the SST. We determined the magnification factor required to match the SST spatial sampling, applied bi-variate spline interpolation, and aligned the IRIS maps to the same FOV as the Ca II 8542 Å SST maps. A similar procedure was applied to the AIA observations to align them with the SST datasets. The final aligned datasets from SST and IRIS, with an effective spatial sampling of ≈0.0591″ pixel$^{-1}$ are shown in Figure 1.

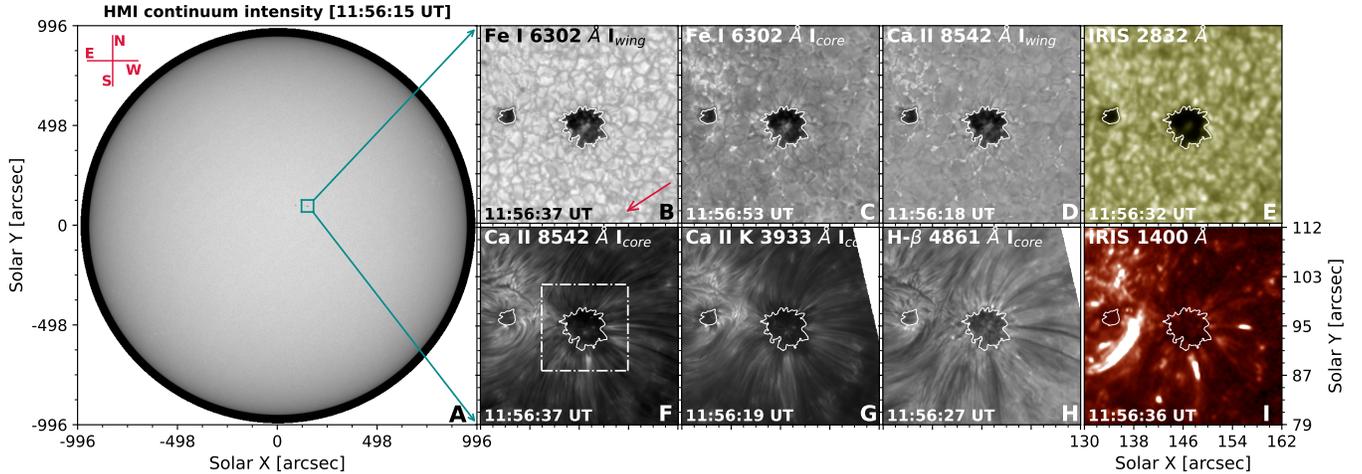

**Figure 1.** Panel A displays the full-disc photospheric continuum intensity image from HMI, with the pores enclosed inside the solid box. Panels B–D and F-H show aligned intensity images in different spectral lines from the SST. Panels E and I show co-aligned UV intensity images from IRIS. White contours outline the pore boundaries. Solar North(N), South(S), East(E), and West(W) are marked on panel A for reference. The arrow on panel B points toward the solar disc centre. The dash-dotted box in Panel F indicates the FOV selected for further analysis.

| Instrument | Lines (Å) | Start-time (UT) | End-time (UT) | Cadence (s) | Spectral position sampled (Å) |
|---|---|---|---|---|---|
| SST | Fe I 6302 | 11:56:37 | 13:14:20 | 37 | -0.3, ±0.26, ±0.22, ±0.18, ±0.14, ±0.1, ±0.06, ±0.02 |
| | Ca II 8542 | 11:56:18 | 13:14:01 | 37 | ±2, ±1.6, ±1.2, ±1, ±-0.8, ±0.6, ±0.5, ±0.4, ±0.3, ±0.2, ±0.1, 0 |
| | Ca II K | 11:56:12 | 12:04:07 | 59 | ±2.08, ±1.82, ±1.56, ±1.36, ±1.17, ±0.97, ±0.78, ±0.59, ±0.52, ±0.45, ±0.39, ±0.33, ±0.26, ±0.19, ±0.13, ±0.07, 0 |
| | H-$\beta$ | 11:56:19 | 12:04:15 | 59 | ±1.82, ±1.56, ±1.3, ±1.04, ±0.78, ±0.52, ±0.46, ±0.39, ±0.33, ±0.26, ±0.2, ±0.13, ±0.07, 0 |
| IRIS | SJI 2832 | 10:11:37 | 12:15:38 | 38 | 2832 |
| | SJI 1400 | 10:11:37 | 12:15:38 | 12.5 | 1400 |
| AIA | (E)UV filters | 11:56:17 | 12:04:12 | 12 | 1600, 1700, 304, 171, 211 |

**Table 1.** Overview of the SST, IRIS, and AIA observations used in the analysis. The spectral positions are mentioned with respect to the line centre of the respective spectral line.

### 3.1. Residual cross-talk correction

Although the SST observations were processed using the existing pipeline, the polarised intensity maps in the Fe I 6302 Å line exhibited residual cross-talk from circular (Stokes V) to linear (Stokes Q and U) polarisation states. The cross-talk appeared in small patches, primarily inside the pore, where the linearly polarised spectra (Q and U) showed a clear trend of the circularly polarised spectra (V). We corrected the Fe I 6302 Å observations for the residual cross-talk before further analysis. For spatial locations with significant polarization degree ($\sqrt{Q^2 + U^2 + V^2} \geq 5\%$), the linear correlation coefficient of the spatial patterns between Stokes Q(U) and V was calculated. We constructed the spatial correlation maps for different values of additional V to Q(U) cross-talk, ramping the V to Q(U) cross-talk value from -30% to 0% and calculated the average correlation value as a function of cross-talk. The average spatial correlation attains its minimum value for the actual crosstalk present in the data (see Figure 6 in Liu et al. (2022)). The linearly polarised maps were corrected for the obtained residual V to Q ($\approx 13\%$) and U($\approx 11\%$) cross-talk before further analysis.

### 3.2. Bisector Analysis

We estimated the line-of-sight (LOS) velocity over ten bisector layers spanning the photosphere and chromo-sphere. For the Ca II 8542 Å line, nine bisector levels were selected, ranging from 10% to 90% of the line depth in uniform steps of 10%, where 90% corresponds to the bisector closest to the line-core. Additionally, one bisector level at 75% of the line depth was selected for the photospheric Fe I 6302 Å line. The bisector levels were chosen to capture LOS velocity variations across multiple layers of the photosphere and chromosphere.

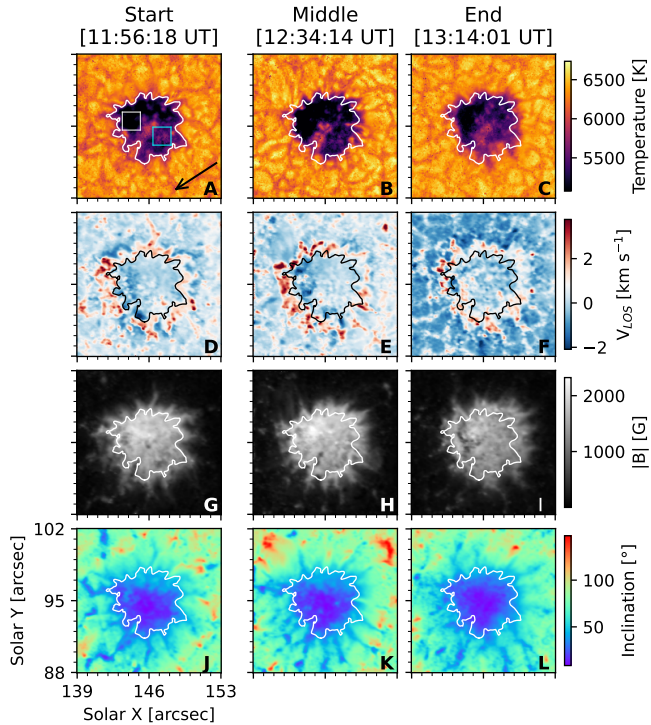

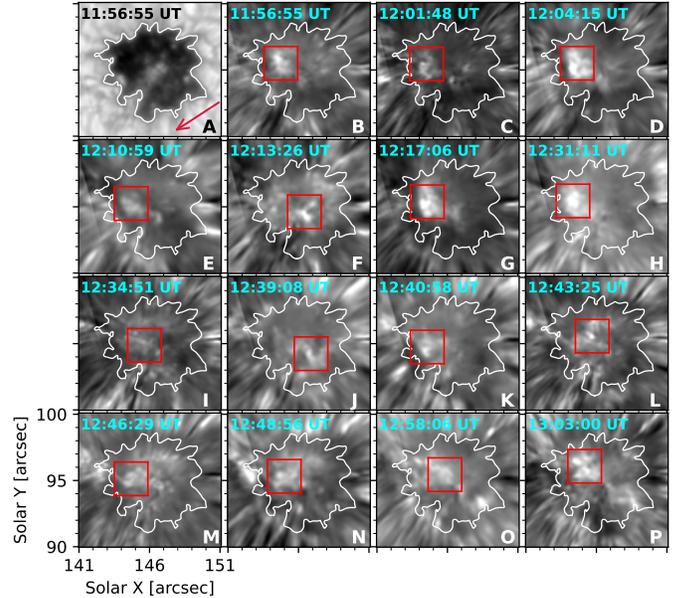

**Figure 3.** Panel A shows the photospheric continuum intensity image in the Fe I 6302 Å line for comparison. Panels B to P present the difference intensity maps at the line centre of the Ca II 8542 Å line during flash events (highlighted with red squares). The arrow on panel A points toward the solar disc centre. A time-series movie of the Ca II 8542 Å line centre maps over 78 minutes of observation, showing multiple flashes inside the pore, is also provided with the manuscript.

**Figure 2.** Panels A–C show the photospheric temperature maps at the start, middle, and end of the observation time window. Panels D–F display the LOS velocity maps, and G–L display magnetic field strength and inclination maps for similar time stamps. The boxes in panel A mark the regions used to compute the average physical parameters for locations with flashes (white box) and without flashes (cyan box). The arrow on panel A points toward the solar disc centre. All maps correspond to the optical depth layer log $\tau \approx -1$.

### 3.3. *Inversion*

The photospheric maps of temperature, LOS velocity, and magnetic field parameters were obtained from inversions of the Fe I 6302 Å spectra (I,Q,U,V) using the Stokes Inversion based on Response functions (SIR[1]; Ruiz Cobo & del Toro Iniesta 1992) code. Out of the two inversion cycles, the first cycle was initialized with the Harvard–Smithsonian Reference Atmosphere

(HSRA; Gingerich et al. 1971), and the resulting atmosphere was used as the initial model for the second cycle. We used 3, 2, and 2 inversion nodes for temperature, LOS velocity, and magnetic field strength (|B|), respectively. Magnetic field inclination, azimuth, and microturbulence used a single node each. We did not provide the specific optical depth layers for node selection and allowed the nodes to be equidistantly spaced over the optical depth range used for inversion, as per the standard algorithm in SIR. The temperature, LOS velocity, and magnetic field parameter maps for the start, middle, and end of the observation window are shown in Figure 2.

The Ca II 8542 Å spectra (I,Q,U,V) were inverted using the Non-LTE Inversion Code using the Lorien Engine (NICOLE[2]; Socas-Navarro et al. 2015). NICOLE performs non–local thermodynamic equilibrium (non-LTE) radiative transfer, synthesis, and inversion of spectral lines, including Zeeman-induced Stokes profiles. It uses a six-level atomic model for the Ca II infrared lines (see Socas-Navarro et al. (2015) for details). We adopted the semi-empirical model C from Fontenla et al. (1993)

---

[1] https://github.com/cdiazbas/SIRcode

[2] https://github.com/hsocasnavarro/NICOLE

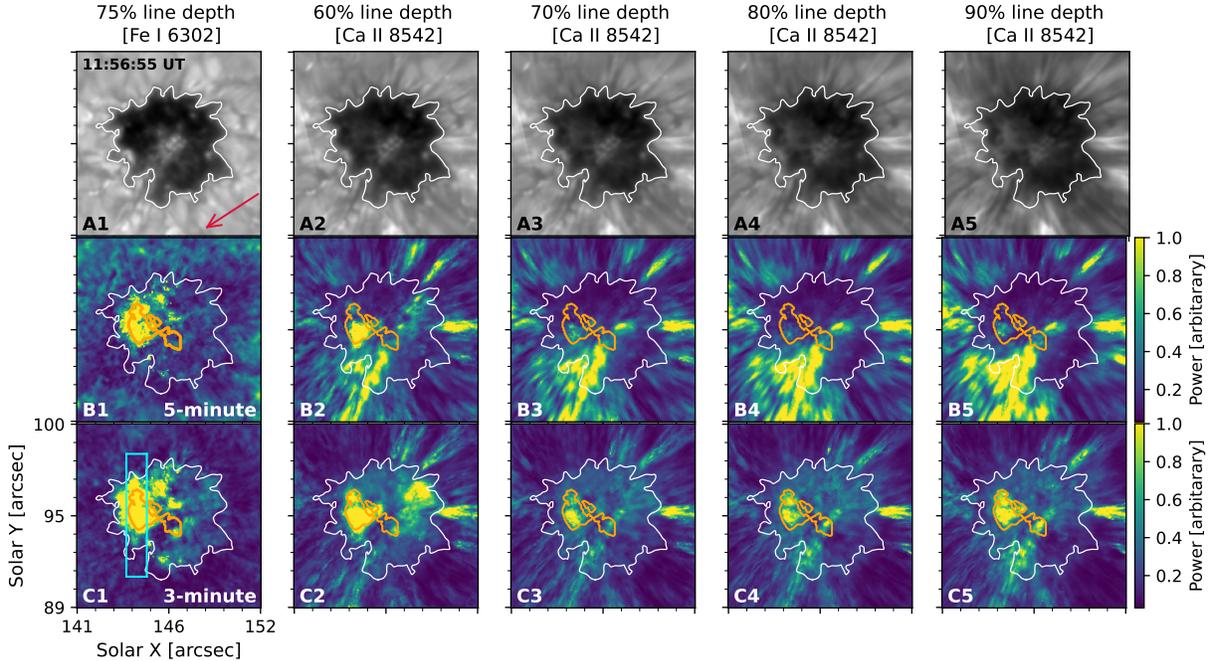

**Figure 4.** Panels A1–A5 show intensity images at different line depth layers of the Fe I 6302 Å and Ca II 8542 Å lines. Panels B1–B5 and C1–C5 show the global wavelet power distribution in and around the pore in the 5-minute and 3-minute bands for similar layers, normalized to the local maximum for each layer. White contours indicate the pore boundary, while the orange contour encloses the regions where flashes occurred frequently. The arrow on panel A1 points toward the solar disc centre. The vertical rectangle in panel C1 marks the FOV used for the inversion of Ca II 8542 Å spectra.

(FALC) as the initial thermal model and performed the inversion using 6 nodes in temperature, 5 nodes in LOS velocity, and 2 nodes each for the vertical magnetic field component ($B_z$) and microturbulence. The location of the nodes in optical depth was again automatically determined by NICOLE. The horizontal magnetic field components ($B_x$ and $B_y$) were assigned one node each due to the noisy spectra in the linear polarisation states (Stokes Q and U). The Ca II 8542 Å spectra were normalized to the Fourier Transform Spectrometer (FTS; Kurucz et al. 1984) spectrum before the inversion. We normalized an average QS spectrum to the FTS and then applied the same normalization coefficient to all Ca II 8542 Å spectra. We utilized temperature, LOS velocity, and $B_z$ stratifications from Ca II 8542 Å inversions in this study.

## 4. RESULTS

Out of the two pores in the final aligned datasets, we analysed the larger pore on the right side of the FOV in this study. The pore had an approximate spatial extent of $8 \times 8 \, \mathrm{arcsec}^2$, centered at x, y = 147″, 95″ from the solar disc center. Multiple flashes were observed inside the pore during the observing period (see Figure 3). The typical area affected by flashes was $3 \times 3 \, \mathrm{arcsec}^2$, appearing every 2-3 minutes, similar to their appearance in sunspots. Many of the flashes exhibit an irregularly shaped internal fine structure. Most (13 out of 15) were located in the left half (x ≤ 145″) of the pore (see Figure 3).

### 4.1. *Flashes and their relation to the photosphere*

Flashes occurred predominantly in the left half of the pore and were largely absent from the right half, with more photospheric structures such as UDs and LB. Some flashes were strong and localized, while others were weak, covering a relatively larger area of the pore. The area/outer boundary of the pore remains pretty much unchanged over the observation spanning more than an hour. The photospheric LB persisted throughout the observing window.

To quantify the differences in physical conditions inside the pore in the photosphere, we selected two regions of $30 \times 30 \, \mathrm{pixel}^2$ within the pore: one located at a flash site (white box in panel A of Figure 2) and the other in a quiet region (cyan box in panel A). The flash-prone locations were co-spatial with relatively cooler areas, having an average temperature approximately 400 K lower than the quiescent pore. The average LOS velocity in the flash region indicated a photospheric upflow of ∼0.5 km s$^{-1}$, opposite to the downflow of ∼0.4 km s$^{-1}$ in the quiescent pore. The average magnetic field strength over the flash region was approximately 250 G stronger than the quiescent pore. Furthermore, the magnetic field lines in flash locations were more inclined ($\sim25°$ with respect to the LOS) compared to those in the quiescent pore ($\sim18°$). The physical conditions mentioned above are for the start of the observations, but the values remained similar at the middle and end of the observing window.

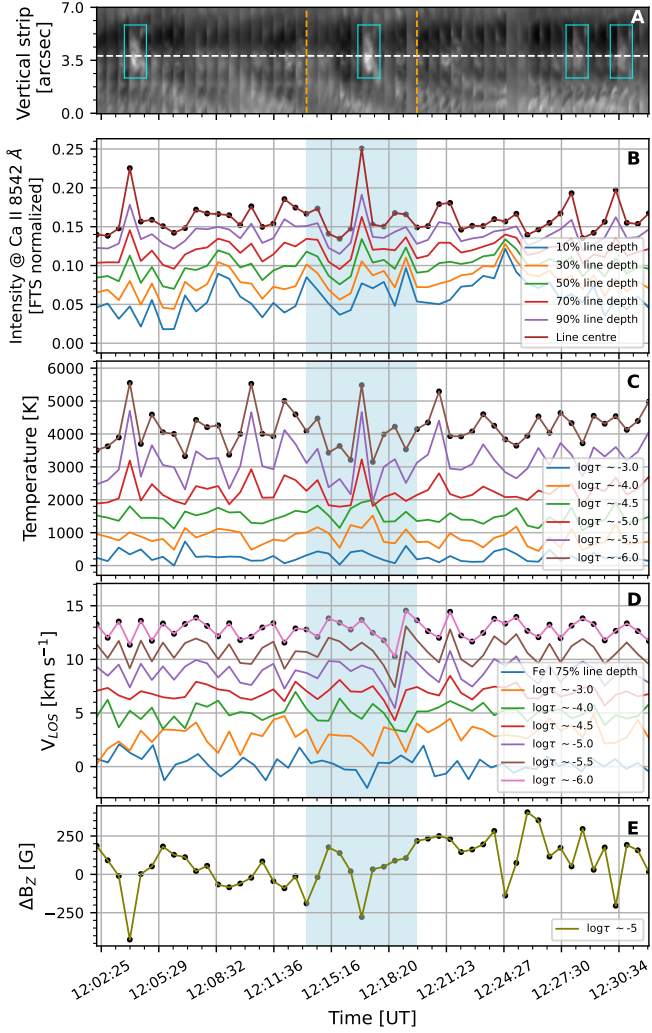

**Figure 5.** Panel A shows the temporal evolution of Ca II 8542 Å line-core intensity maps over the strip indicated in panel C1 of Figure 4. Cyan rectangles highlight flashes. Panel B presents the intensity variation at different line depths layers of the Ca II 8542 Å line, averaged over the white dashed slit in panel A at each time step. Panels C and D display temperature and LOS velocity variations for different optical depth layers averaged over the same slit. Intensity, temperature, and LOS velocity variations for each layer are plotted after removal of the mean value and are added with a constant offset for consecutive higher layers. Panel E shows a change in $B_z$ variations for the same time window. The region highlighted with the light blue background is similar to that enclosed by dashed vertical lines on panel A.

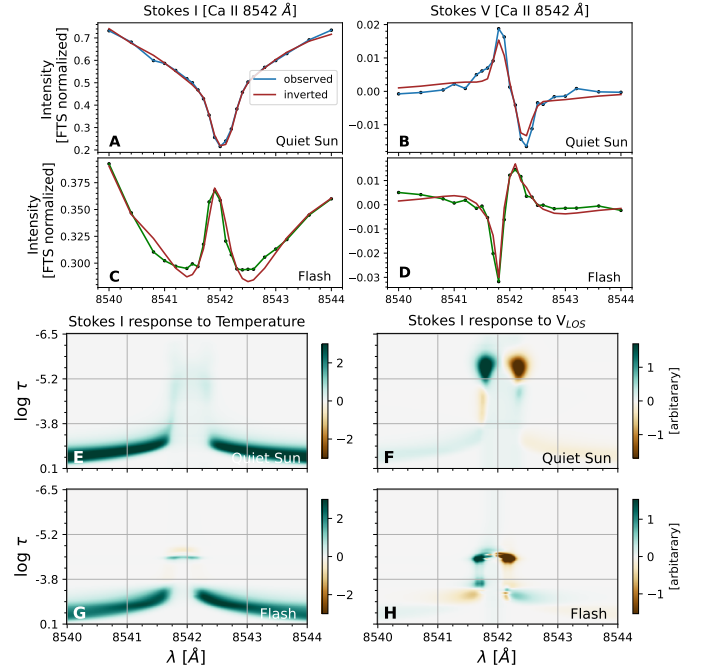

**Figure 6.** Panels A and B show the observed and inverted Stokes I and V spectra for a quiet-Sun location (point 1 in panel A of Figure 7). Panels C and D present similar spectra for a flash location (point 2 in panel A of Figure 7). Panels E and F display the response functions of Stokes I to temperature and LOS velocity, respectively, for the quiet-Sun location. Panels G and H show the corresponding response functions for the flash location.

### 4.2. Power distribution and its relation to Flashes

We examined the spatial distribution of oscillatory power in and around the pore to investigate how the occurrence of flashes is related to the underlying power distribution. The power was calculated in the 3- and 5-minute bands (averaged over ±0.1 minutes) using the global wavelet power spectrum (Torrence & Compo 1998) of the LOS velocities. Four bisector levels at 60%, 70%, 80%, and 90% line depth of the Ca II 8542 Å line, and one at 75% line depth of the Fe I 6302 Å line, were selected to characterise the power distribution across multiple layers of the photosphere and chromosphere. The intensity at the corresponding line depths is shown in panels A1–A5 of Figure 4. The intensity map at 75% line depth Fe I 6302 Å line (panel A1) samples the photosphere showing granulation, UDs and the LB inside the pore, whereas the Ca II 8542 Å map at 90% line depth (panel A5) captures chromospheric structures such as superpenumbral fibrils along with a flash inside the pore.

In the photosphere, the 3-minute and 5-minute power is predominantly confined within the pore, with a slightly enhanced 5-minute power outside the pore in granulation areas as well (see panels B1 and C1 of Figure 4). Strongly enhanced 5-minute power is co-spatial with the flash location (enclosed with orange countours) in the photosphere and extends out from the pore for upper layers of the atmosphere, seemingly along chromospheric fibrils (see panels B2 to B5). On the contrary, the 3-minute power continues to be enhanced at the flash location in upper layers as well (see panels C2 to C5). The LB shows an enhancement in the 5-minute power for upper layers of the atmosphere (see panels B1–B5 at x,y≈ 146″,95″). The locations with increased 5-minute power near the LB stay inconspicuous in the 3-minute power maps.

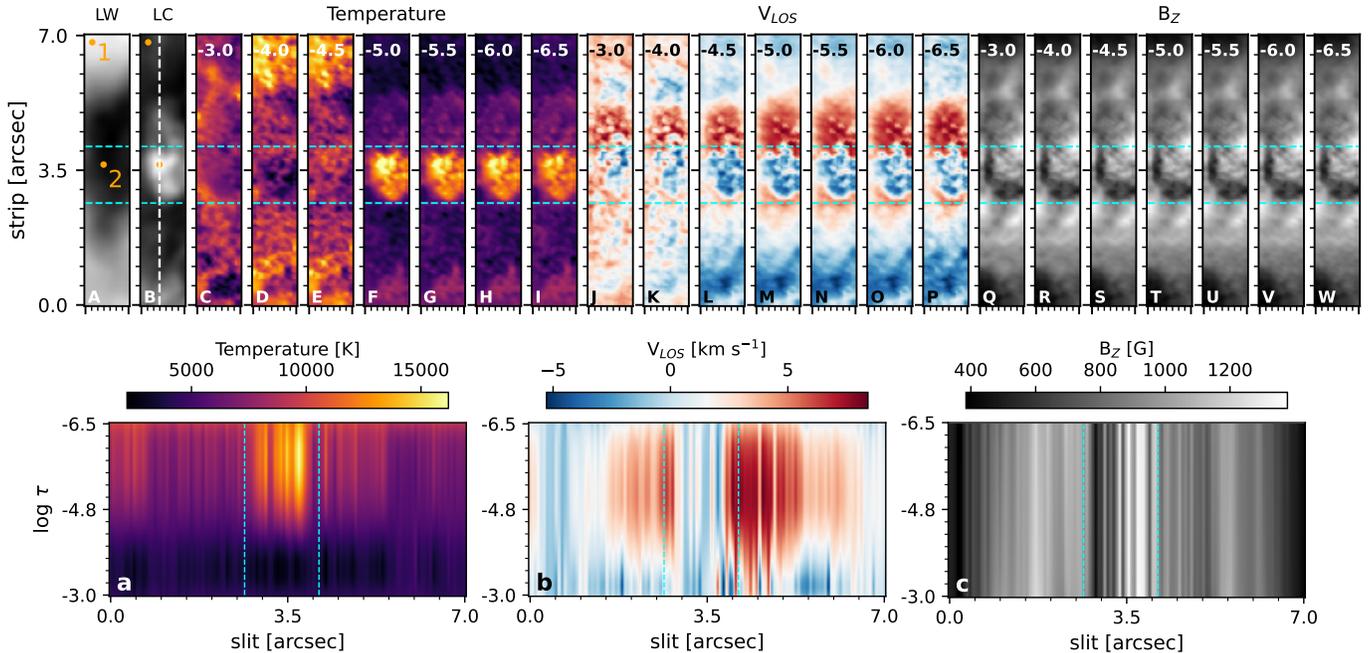

**Figure 7.** Panels A and B show the Fe I 6302 Å line-wing and Ca II 8542 Å line-core intensity maps over the strip for a strong flash appearing close to 12:17 UT. Points 1 and 2 mark the spatial locations for which spectra and response functions are plotted in Figure 6. Panels C–I show the temperature stratification, Panels J–P show the LOS velocity stratification, and Panels Q–W show the $B_z$ stratification over the strip for different optical-depth layers. Panels a, b, and c display the vertical cross-sections of temperature, LOS velocity, and $B_z$ stratification along the cut along the y-axis on Panel B. Cyan dashed lines on the different panels enclose the flash.

### 4.3. *Flashes in the Chromosphere*

Co-spatial with flash-prone locations inside the pore with enhanced power in 3- and 5-minute bands, we selected a vertical strip (20×120 pixels²) for further analysis (panel C1 of Figure 4). The strip covered a portion of the quiet Sun outside the pore as well. The temporal evolution of the Ca II 8542 Å line-centre intensity over the strip is shown in panel A of Figure 5. Multiple flashes can be seen in the Ca II 8542 Å time series (highlighted with cyan rectangles in panel A), with a particularly strong flash occurring near 12:17 UT. Mul-tiple running disturbances, triggered by flashes inside the pore, can also be seen.

To further investigate the flash-associated intensity variations across different layers of photosphere and chromosphere, we plotted the temporal variation of the intensity averaged along the slit (white dashed line in panel A) near the flash location. When flashes occur, distinct intensity enhancements are observed up to the 50% line-depth layer of the Ca II 8542 Å line, but not below (panel B). Flashes produce a temperature enhancement, clearly detected from $\log \tau \sim -5$ (i.e $\tau$ represents optical depth at 5000 Å) upwards, with the enhancement ranging from ∼500 K at $\log \tau \sim -5$ to ∼2500 K at $\log \tau \sim -6$ with respect to the quiescent phase (panel C). For the strong flash at 12:17 UT, with a time delay of ∼2 minutes, the LOS velocity shows a drop of up to 5 km s⁻¹ at $\log \tau \sim -5$ following the flash (panel D). The velocity drop is observed in both higher and lower layers with no signature in the photospheric Fe I 6302 Å velocity variations. The weaker flashes at UT 12:28 and 12:30 show a similar behavior but of much smaller amplitude.

Lower atmospheric layers do not show clear temperature or velocity signatures associated with flashes. Each flash event is accompanied by a decrease of $B_z$ by ∼100–200 G compared to the quiescent phase (panel E), where the $B_z$ provided by NICOLE should be close to the magnetic field component along the local vertical because of the small ($\approx 15$ °) heliocentric angle.

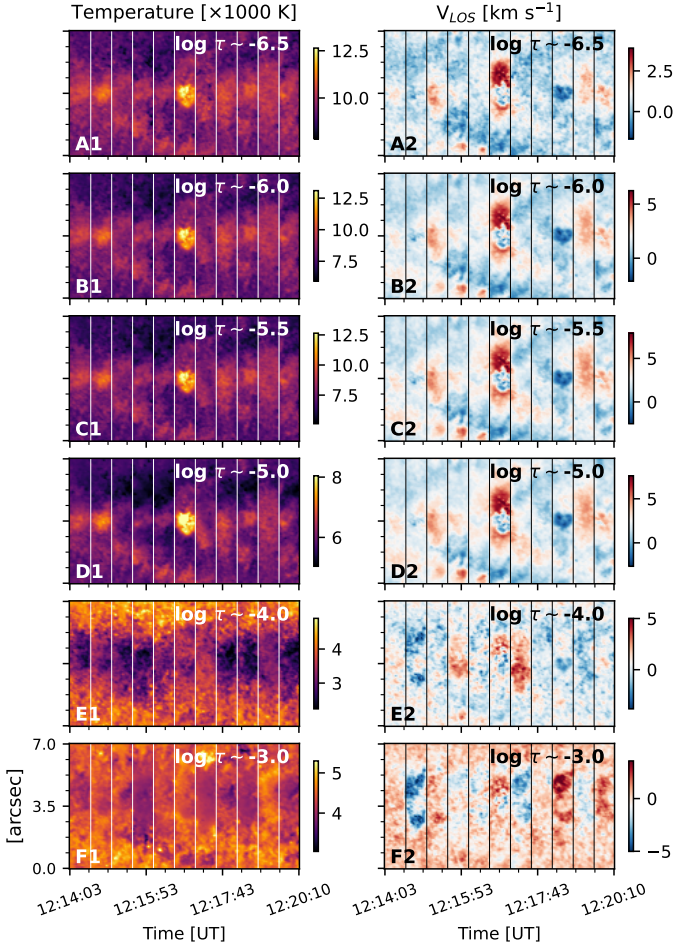

**Figure 8.** Left panels (A1–F1) and right panels (A2–F2) show the temporal evolution of 2D stratifications of temperature and LOS velocity over different optical-depth layers close to the strong flash appearing at 12:17 UT.

We plotted the observed and inverted spectra for a flash and a quiet-Sun location outside the pore in panels A–D of Figure 6. The spatial locations were selected from the intensity maps of the strong flash at $\sim$12:17 UT (points 1 and 2 in panels A and B of Figure 7). The observed spectra are well reproduced by the inversion at both locations.

Additionally, we computed the response functions of Stokes I to temperature and LOS velocity using the multi-atom non-LTE Stockholm inversion Code (STiC; de la Cruz Rodríguez et al. 2019) to determine the optical depth ranges where Stokes I is more sensitive to perturbations (bottom panels of Figure 6). For both locations, we used the NICOLE-inverted spectra and model atmospheres to estimate the response functions

(panels E–H of Figure 6). In the quiet Sun, Stokes I is more responsive to temperature perturbations below $\log \tau \sim -4$ and only weakly responsive between $\log \tau \sim -4$ and $-6$. During a flash, the peak temperature sensitivity shifts to lower optical depth layers. The LOS velocity response of Stokes I is strong between $\log \tau \sim -5$ and $-6.5$ in the quiet Sun, which shifts toward the lower optical depth layers at $\log \tau \sim -4.5$ during a flash.

#### 4.3.1. *For the strong flash at 12:17 UT*

We inverted the Ca II 8542 Å spectra over the selected strip to retrieve the 2D maps of temperature, LOS velocity, and $B_z$ stratifications for the strong flash observed near 12:17 UT (see panel B of Figure 7). A distinct temperature enhancement, co-spatial with the flash location, can be seen from $\log \tau \approx -5$ upwards (panels C–I), without a significant increase in lower layers. At $\log \tau \approx -5$, the flash location shows an upflow surrounded by downflows in its vicinity; that weakens both in higher and lower layers (panels J–P).

Complimentarily, the stratification of temperature, LOS velocity, and $B_z$ along the cut along the y-axis marked in panel B is displayed in panels a–c. The flash-induced temperature enhancement, upflow, and $B_z$ drop are highlighted within the dashed lines. The flash stands out in temperature and LOS velocity maps compared to the surrounding region. Relative to the background pore, the flash produced a temperature enhancement of $\sim 2500$ K at $\log \tau \approx -6$, diminishing in lower layers. The atmosphere between $\log \tau \approx -3$ and $-6.5$ exhibits upflows at the flash location, surrounded by downflows. The $B_z$ decreases by $\sim 200$ G at the flash location compared to the surrounding pore.

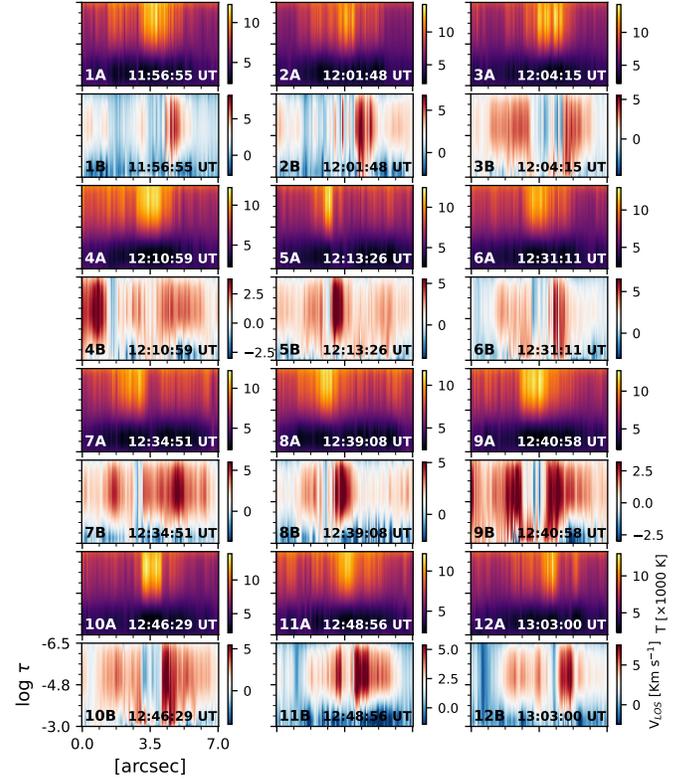

**Figure 9.** Panels 1A–12A and 1B–12B show the vertical slice of temperature and LOS velocity stratifications over different optical-depth layers for twelve distinct flashes.

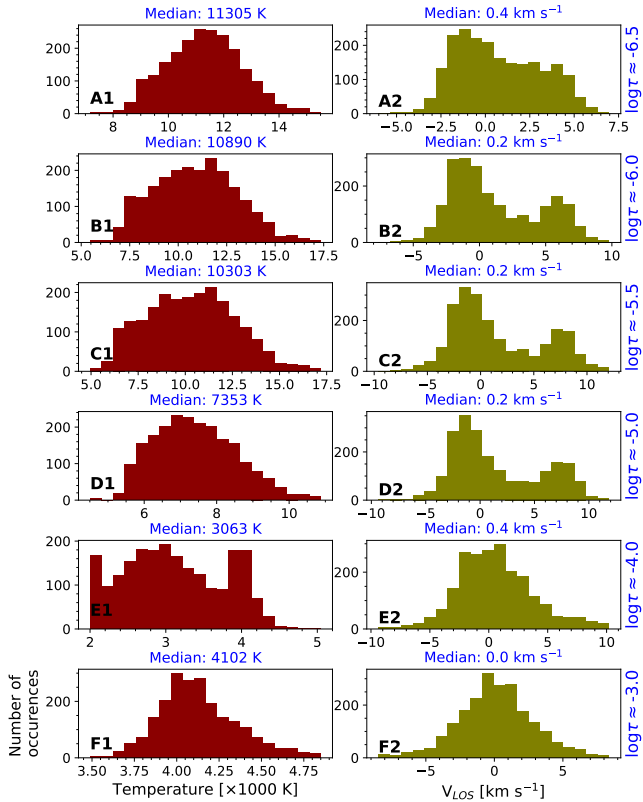

**Figure 10.** Statistical distribution of temperature (panels A1–F1, left) and LOS velocity (panels A2–F2, right) across different optical-depth layers for all spatial locations associated with flashes.

We plotted the temporal evolution of temperature and LOS velocity maps for the strong flash ($\sim$12:17 UT) at the fastest 37-second cadence of the observations (see Figure 8). The temperature enhancement is evident from $\log \tau \approx -5$ to $-6.5$, weakening with height, and is absent in the pre- and post-flash phases. The LOS velocity variations are more complex. During the flash, a central upflow is surrounded by downflowing flanks. 37 s after the flash, the atmosphere is downflowing at $\log \tau \approx -4$, sandwiched between upflowing atmosphere (at $\log \tau \approx -3$ and above $\log \tau \approx 5$). As time progresses, $\log \tau \approx -3$ shows strong downflows, whereas the layers above are still upflowing. Temperature enhancements are seen to propagate away from the flash site (see panels A1–D1), co-spatially with strong upflows (panels A2–D2). The 2D maps of temperature and LOS velocities before and after the flash appear almost disconnected from the flash map, indicating that the 37-second cadence is insufficient to capture the temporal evolution of different physical quantities during the flash.

### 4.3.2. *For twelve different flashes*

The LOS velocity stratification for the strong flash ($\sim$12:17 UT) shows the central upflow ($1''$–$2''$), surrounded by downflowing flanks in its vicinity (see panel b of Figure 7). The downflowing flanks are asymmetric, extending over a slightly larger area than the central upflow. To determine whether the velocity pattern is a general characteristic or if it can reverse in other cases, we selected twelve flashes within the pore that appeared at different times. For each flash, a cut along the y-axis passing through its centre was chosen, and the Ca II 8542 spectra along the strip were inverted to retrieve the temperature (panels 1A–15A) and LOS velocity stratification (panels 1B–15B). In all twelve cases, the flash centres exhibit an upflow, surrounded by downflowing flanks in their vicinity, indicating that the LOS velocity pattern observed for the strong flash is a consistent feature of other flashes as well.

### 4.3.3. *Statistical properties of all flashes*

Complementing the 2D maps, we analysed the stratifications of temperature and LOS velocity at all spatial locations associated with flashes (enclosed by the orange contour in Figure 4). A location was classified as flash-associated if the Ca II 8542 Å line-centre intensity exceeded 40% of the average quiet sun (over $100\times100$ pixel$^2$) Ca II 8542 Å line-wing ($\Delta \lambda \approx$ -2 Å) intensity, yielding 2164 profiles over the observation window. The temperature enhancement distributions are skewed with an extended tail towards higher temperatures between $\log \tau \approx -5$ and $\log \tau \approx -6.5$, where flashes are more prominent (see panels A1–F1 in Figure 10). The LOS velocity distribution is bimodal, with $\approx$52% of profiles showing downflows at $\log \tau \approx -5$ (see panels A2–F2). The bimodal character of the LOS velocity distribution is also evident in layers between $\log \tau \approx -4.5$ and $\log \tau \approx -6.5$, whereas at $\log \tau \approx -3$ and -4, the distribution is roughly single Gaussian.

### 4.3.4. *Flash in relation to running (penumbral) waves*

In the majority of cases, the flashes inside the pore were followed by radially outward propagating running waves (RWs), very similar to the running penumbral waves observed around sunspots in the chromosphere. Since the pore does not have a penumbra, we re-frain from calling them running penumbral waves. The Ca II 8542 Å line-center intensity stack over the strip (panel A of Figure 5) shows multiple RWs (arc-shaped disturbances), each of which can be linked one-to-one with a flash inside the pore, as also seen in the LOS velocity stacks of Figure 8 (see panels A2–F2).

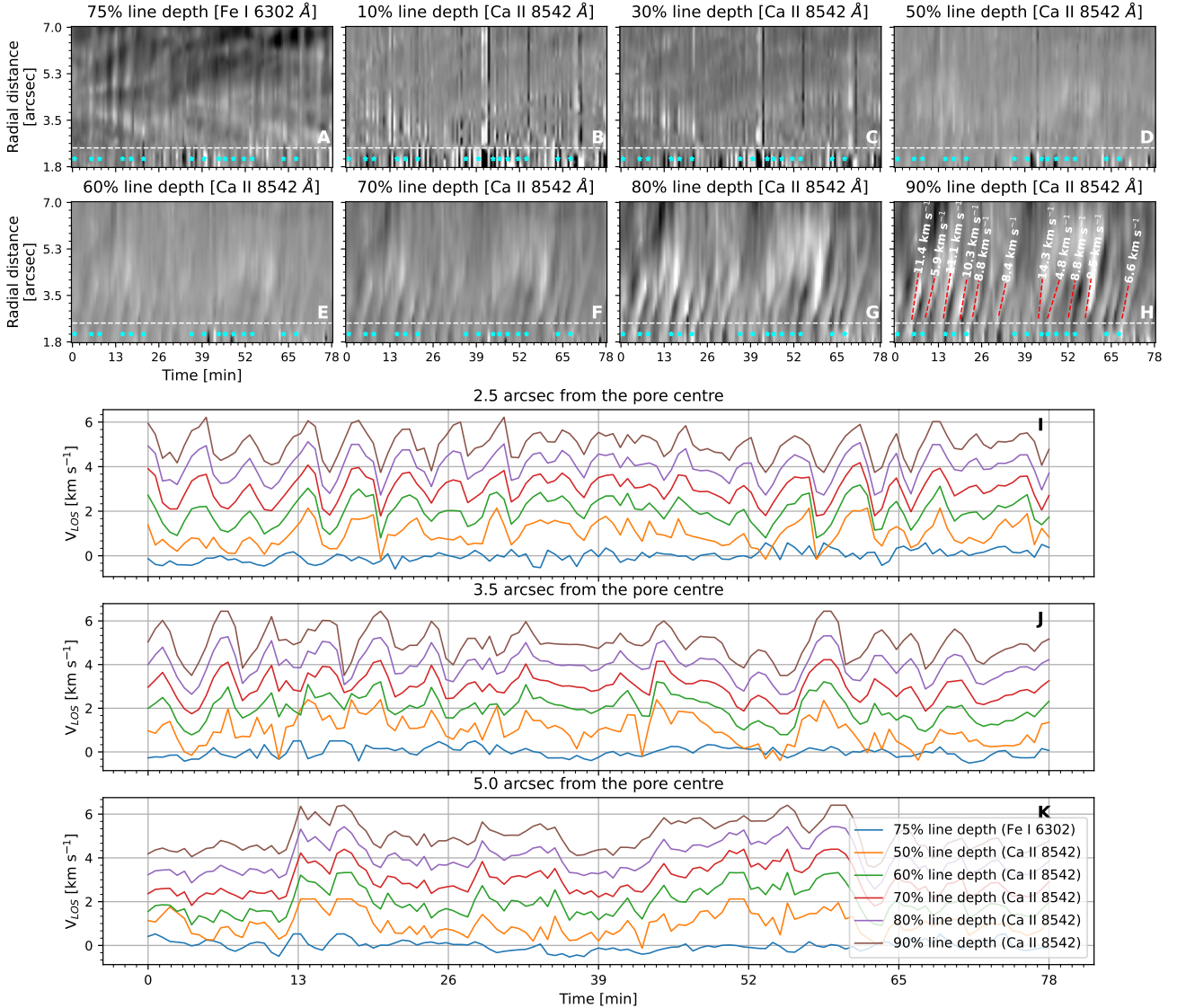

**Figure 11.** Temporal evolution of the azimuthally averaged radial variation of LOS velocity for different line-depth layers in Fe I 6302 Å (panel A) and Ca II 8542 Å (panels B–H) lines. The radial variations were averaged over ±5°, centered at 210° in azimuth around the pore. Multiple RWs, along with their radial propagation speeds, are marked in panel H. Horizontal dashed lines in the panels mark the pore boundary. Cyan stars inside the pore boundary indicate the time instances at which flashes appeared. Panels I, J, and K display the temporal evolution of LOS velocities associated with RWs across different line depth layers and different radial distances, plotted after removal of the mean value and added with a constant offset of 1 km s⁻¹ for consecutive higher layers.

The temporal evolution of the LOS velocity, averaged over an angular cone of ±5°, centered at 210° at different line-depth layers of the Fe I 6302 Å and Ca II 8542 Å lines was plotted to investigate the behavior of RWs across different layers (see Figure 11). Multiple radially propagating RWs are visible as bright and dark streaks in the LOS velocity stacks (see panels G and H). The RWs can be seen clearly across multiple layers spanning from the 90% line-depth layer (panel H) to 50% line-depth layer (panel D), but not below. The photospheric velocity stack (panel A) also does not show an explicit presence of RWs in it. The radial extent of RWs is confined to within 2″ from the pore boundary. We estimated the propagation speed of the RWs using a linear fit to the position-time diagram of the propagating velocity streaks. The red dashed line on different RWs in panel H shows the linear fit with the propagation speed (5-15 km s$^{-1}$) of the respective RW marked on it.

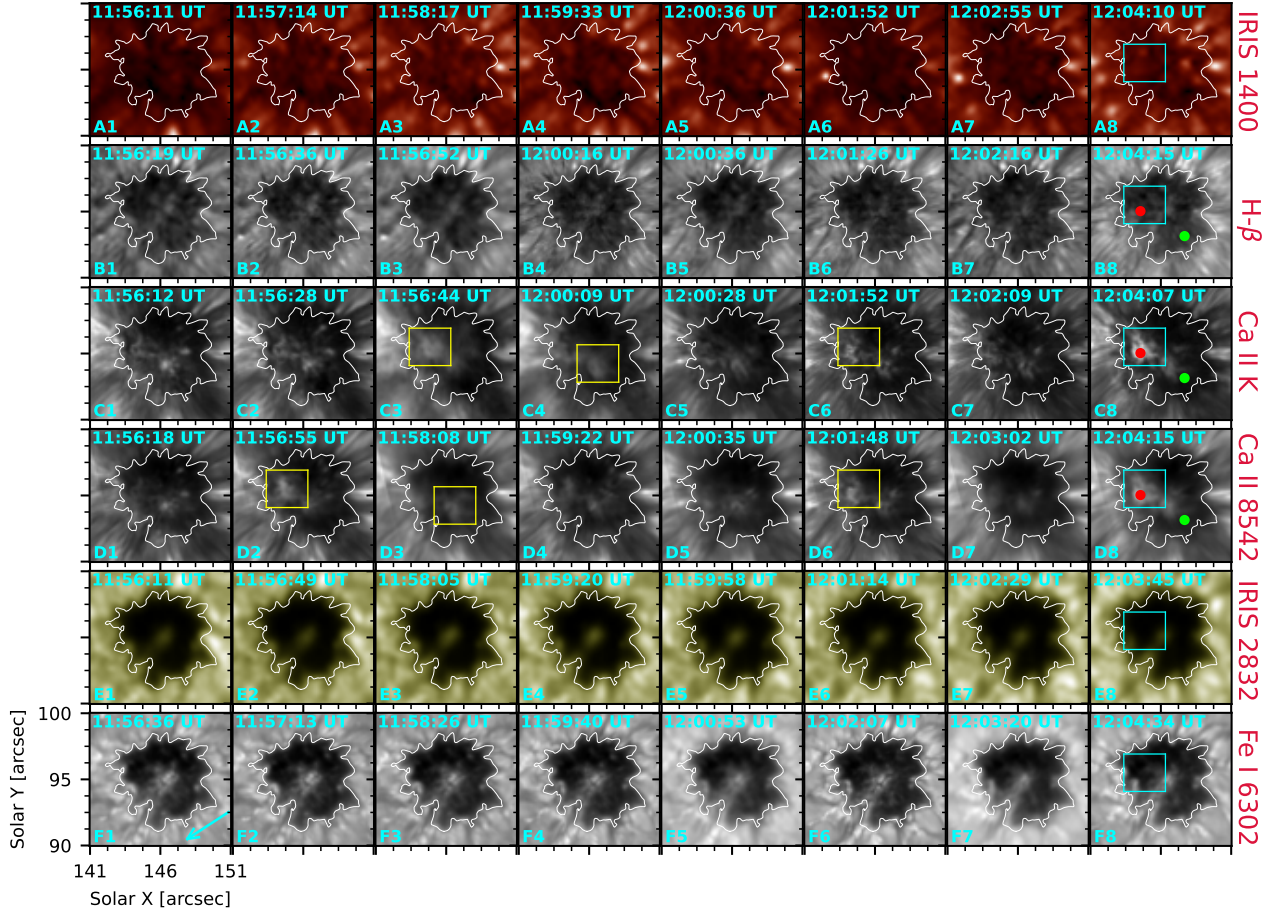

**Figure 12.** Temporal evolution of the intensity maps in different diagnostics, arranged from top to bottom in approximate decreasing formation-height order. Cyan boxes in panels A8–F8 enclose a flash that is prominent in the Ca II (8542 & K) lines, while yellow boxes highlight flashes appearing earlier. The red (green) dots mark the spatial location used for plotting the spectra and intensity evolution in Figures 13 and 14. The white contour marks the pore boundary. The arrow on panel F1 points toward the solar disc centre.

We plotted the temporal evolution of the LOS velocity perturbations associated with RWs at three different radial distances from the pore (see panels I–K of Figure 11). Close to the pore boundary (2.5″ from the pore center), at the 90% line depth layer of Ca II 8542, the velocity amplitudes associated with RWs were of the order of 1 km s$^{-1}$ that decreased in the lower layers (see panel I). As the RWs propagated radially outward, the amplitude increased from ∼1 km s$^{-1}$ to ∼2 km s$^{-1}$ within 1″ at the 90% line depth layer due to damping (see panel J) and almost disappeared within the next 1.5″ (see panel K). Even though the 2D stacks of LOS velocities do not show RWs in the upper photosphere (panels A and B of Figure 11), the line plots at fixed radial distances do show the LOS velocity variations in the upper photosphere, similar to the velocity variations associated with the RWs in the chromospheric layers (see variations at ∼13 and 60 minutes in panels I–K).

### 4.4. Flashes at the transition region and above

The LOS velocity structuring of a flash features a central upflow in the chromosphere, which could potentially propagate into the transition region and above. We investigated their possible signature across different layers (see Figure 12). A flash occurring close to 12:04:15 UT is highlighted by a cyan box in the rightmost panels (A8–F8). The flash is clearly visible in the Ca II 8542 Å and Ca II K intensity maps, but appears less prominent in H$\beta$ (panel B1-8). In contrast, the IRIS 1400 Å intensity maps (panels A1-8), which sample transition-region plasma, show no distinct signature of the flash, also not at a later moment. We also investigated intensity maps in EUV channels of AIA (not shown), without any explicit signature of flashes in the solar corona. Similarly, lower-chromospheric intensity maps such as IRIS 2832 (panel E1-8), along with the line-centre intensity map of the photospheric Fe I 6302 Å line (panel F1-8), do not exhibit a distinct flash signature. Other flashes occurring earlier are highlighted inside the yellow boxes in Figure 12. These flashes, too, are not distinctly present in intensity maps of the photosphere or the transition region/corona, nor at later times, indicating the localization of flashes in the chromospheric layers only.

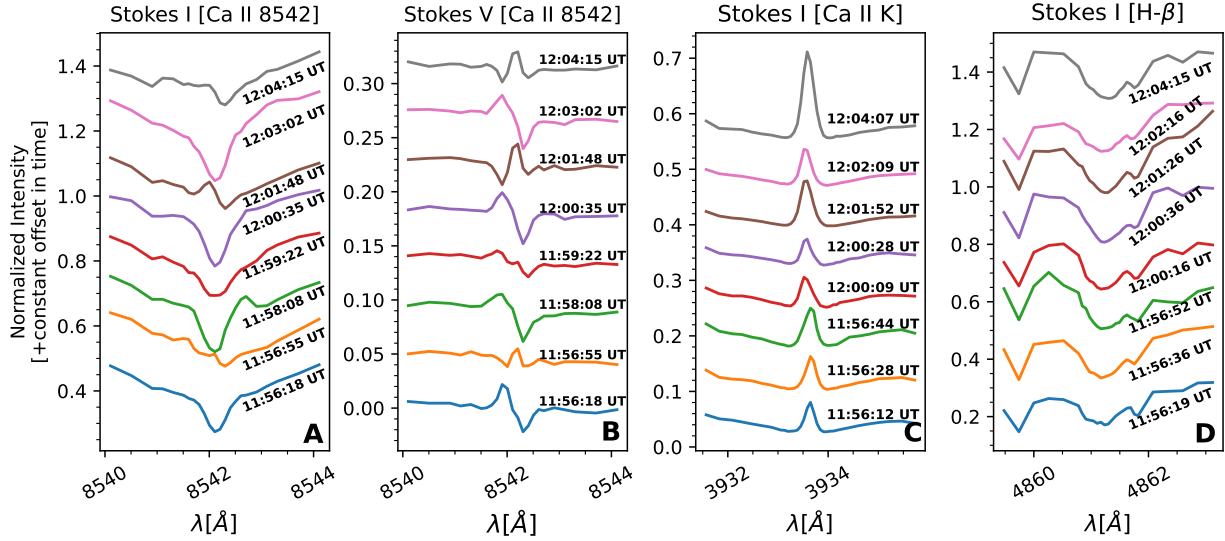

**Figure 13.** Temporal evolution of the spectra (Stokes I and V) in the Ca II 8542 Å, Ca II K, and H-β lines at the flash location (red dot in panels B8–D8 of Figure 12).

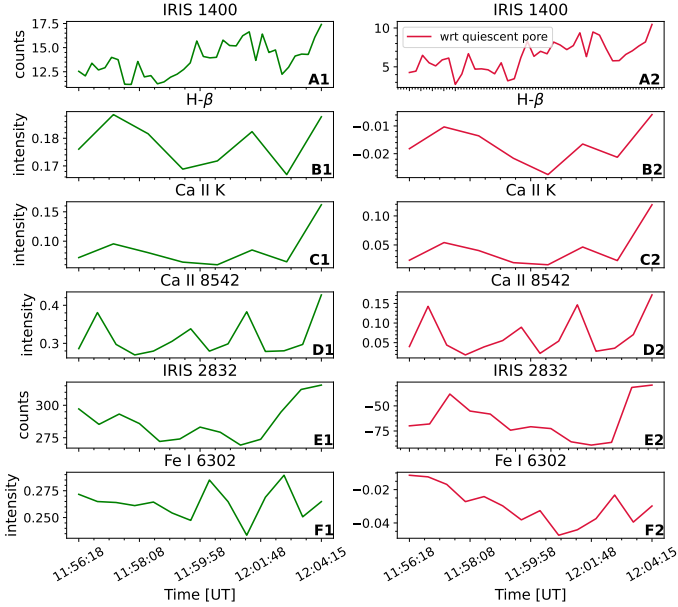

**Figure 14.** Left panels (A1–F1) show the temporal variation of intensity averaged over a $3\times3$ pixel$^2$ box near the flash location (red dot in panels B8–D8 of Figure 12). Right panels (A2–F2) show the corresponding variations relative to the quiescent pore (green dot in panels B8–D8 of Figure 12).

### 4.4.1. Spectral and temporal variations

Additionally, we investigated the spectral response of a flash in Ca II 8542 Å, Ca II K, and Hβ lines (see Figure 13). During a flash (e.g., at 11:56:55 UT, 12:01:48 UT, and 12:04:15 UT), the Stokes I spectrum of Ca II 8542 Å becomes shallower, with an emission feature appearing near the line center. The Stokes V profile in Ca II 8542 Å reacts to the emission close to the line center, leading to an apparent polarity reversal during strong flashes. Both Stokes I and V profiles return to their quiescent shapes before and after the flash. The Ca II K Stokes I spectra generally show a persistent line-center emission feature (panel C of Figure 13), which becomes enhanced during flashes (12:01:48 UT, and 12:04:15 UT), producing distinct localized brightenings in the intensity maps. In contrast, although Hβ is also a chromospheric spectral line, its spectra do not display a spectrally localized brightening comparable to the Ca II 8542 Å or Ca II K lines. Instead, the flash produces an intensity enhancement across the entire Hβ line, shifting the overall intensity level upward during the flash, and returning to normal shape in pre- and post-flash scenarios (e.g., at 11:56:36 UT, 12:01:26 UT, and 12:04:15 UT in panel D of Figure 13).

To quantify the flash-associated intensity enhancement in different spectral lines and filters, we plotted the temporal evolution of intensities averaged over a fixed wavelength range in different lines, averaged over the flash location ($3\times3$ pixel$^2$) (see Figure 14). For each time step, the spectral averaging was done over ±25 mÅ near the line center of the photospheric Fe I 6302 Å line, and over ±100 mÅ near the line center of the chromospheric Ca II 8542 Å, Ca II K, and Hβ lines.

Co-temporal with flash appearance (~11:56:55 UT, 12:01:48 UT, and 12:04:15 UT), distinct intensity peaks can be seen in Ca II K (panels C1–2), Ca II 8542 Å (panels D1–2), and in H-$\beta$ (panels B1–B2). Similar to the spectral response, the temporal response in the Ca II (Ca II 8542 and K lines) is more localised in time compared to H-$\beta$. The transition region (IRIS 1400; panels A1–2) or the photospheric layers (IRIS 2832 Å panels E1–2 and Fe I 6302 Å panels F1–2) do not show such variations corresponding to the flashes. We also examined the intensity variation relative to a quiescent location within the pore. The relative variations were mostly similar to the variation at the flash location that appeared prominently in the chromospheric lines.

The spectral and temporal variations we observed indicate the confinement of flashes in the chromosphere, with no significant impact on either the lower (photosphere) or the higher (transition region and corona) layers.

## 5. DISCUSSION

The observed pore exhibited repeated flash activity over the observation window, consistent with earlier reports of persistent flash occurrence in sunspots (Rouppe van der Voort et al. 2003; Sych & Wang 2018). The presence of a light bridge throughout the observation suggests a locally complex magnetic field structuring, which may influence the spatial localisation and recurrence of flashes (Socas-Navarro et al. 2009). The repeated flashes over a similar region inside the pore indicate that flashes are not random events but are linked to persistent wave-driven processes within the pore atmosphere.

### 5.1. *Flashes in relation to the photosphere*

The occurrence of the majority of flashes, co-spatially with cooler regions with stronger and more inclined magnetic field lines in the photosphere, is consistent with observations of Yurchyshyn et al. (2020) and Koyama & Shimizu (2024). Such conditions favour the upward propagation of magneto-acoustic waves along inclined fields, enabling shock formation in the chromosphere (Centeno et al. 2006; Bard & Carlsson 2010). The absence of UFs in areas with UDs and the LB suggests that small-scale magnetic and thermodynamic structuring can inhibit their development (Socas-Navarro et al. 2009). The photospheric Dopplergram shows small-scale patches of downflows just outside the pore boundary in line with Keil et al. (1999); Hirzberger (2003); Cameron et al. (2007) and Sobotka et al. (2012).

The 3- and 5-minute power enhancements near flash-prone locations in the photosphere and the residual 3-minute power persisting up to the chromosphere support the role of these waves in driving flash shocks. These results indicate that both the local photospheric magnetic and velocity environment and the spatial structuring of oscillatory power control the occurrence and distribution of flashes (Rouppe van der Voort et al. 2003; Sych & Nakariakov 2014). The fanning out of the 5-minute power outside the pore with increasing height in the atmosphere is in line with observations of Löhner-Böttcher & Bello González (2015) in sunspots.

### 5.2. *Flashes in the chromosphere*

The persistence of 3-minute power at flash-prone locations across different layers of the photosphere and chromosphere is consistent with earlier findings of flashes preferentially occurring in areas where magneto-acoustic waves are amplified and steepen into shocks in the chromosphere (Beckers & Tallant 1969; Socas-Navarro et al. 2000; Centeno et al. 2006). We tracked flashes in intensity down to the 50% line depth of Ca II 8542, but not below, supporting the view that flashes are predominantly chromospheric phenomena whose intensity enhancements are caused by localised temperature increases at optical depth layers of log $\tau \sim$ -5 (de la Cruz Rodríguez et al. 2013; Henriques et al. 2017).

We observed the temperature enhancements during flashes to vary from ~500 K at log $\tau \approx$ −5 to ~2500 K at log $\tau \approx$ −6, consistent with shock heating models (Bard & Carlsson 2010; Felipe et al. 2014). The LOS velocity pattern was dominated by a central upflow with downflowing flanks, a configuration suggestive of shock-induced mass redistribution (Felipe et al. 2010; Houston et al. 2020). Associated downflows of ~8 km s$^{-1}$ at log $\tau \approx$ −5, weakening in both upper and lower layers, are in agreement with earlier reports of complex flow patterns during flashes (Socas-Navarro et al. 2001; Yurchyshyn et al. 2014). Importantly, a distinct magnetic signature was observed, where B$_z$ decreases by 100-200 G during flash events, echoing prior observations of transient magnetic field weakening likely linked to opacity effects and shock-induced changes in line formation heights (de la Cruz Rodríguez et al. 2013; Houston et al. 2018).

Response function (RF) analysis shows that during the quiescent phase, Stokes I in Ca II 8542 Å is sensitive to temperature perturbations up to log $\tau \approx$ −6.5, along with a strong photospheric sensitivity. The LOS velocity sensitivity during the quiescence phase peaks between log $\tau \approx$ −5 and −6.5. However, during flashes, both the temperature and LOS velocity sensitivities shift to lower optical depth layers, consistent with earlier studies demonstrating how shocks and flares modify the height of formation and RF distribution of chromospheric lines (de la Cruz Rodríguez & Socas-Navarro 2012; Kuridze et al. 2017; Yadav et al. 2021).

A statistical analysis of 2164 flash-associated profiles confirms that the upflow–downflow configuration seen in the strong flash is typical for most events, aligning with earlier reports of a central upward shock front surrounded by return flows (Yurchyshyn et al. 2020).

The observation of multiple RWs with a one-to-one correspondence with flashes in the majority of cases is notable. The measured apparent lateral propagation speed of 5-15 km s$^{-1}$ is in the expected range for R(P)Ws in sunspot chromospheres (Bloomfield et al. 2007; Yuan et al. 2014). The radial propagation speeds we found were smaller than those found by Löhner-Böttcher & Bello González (2015) in a sunspot (∼37 km s$^{-1}$ in chromosphere, ∼51 km s$^{-1}$ in photosphere). We also observed the steepening of the RWs during radial propagation within the first few arcsecs from the pore boundary. The temporal and spatial coupling agrees with the interpretation that flashes and RWs are manifestations of the same upward-propagating slow magnetoacoustic wave train, with the former representing the shock formation stage (Jess et al. 2013; Krishna Prasad et al. 2015).

Overall, the results support the picture that flashes in pores are signatures of upward-propagating magnetoacoustic shocks that locally heat the chromosphere, modify LOS velocities in a structured manner, and induce transient reductions in the apparent magnetic field strength, with their occurrence intimately linked to the initiation and dynamics of the RWs.

### 5.3. Flashes in transition region and above

The absence of clear flash signatures in the transition region (TR) and coronal diagnostics suggests that the flashes inside the pore were confined primarily to the chromospheric heights. The confinement can also be seen in the flash response to spectra. Strong intensity enhancements observed in the Ca II 8542 Å and Ca II K spectra, with localized emission peaks near the line center, are consistent with earlier observations of chromospheric heating during flashes (Socas-Navarro et al. 2000; Beck et al. 2008, 2013; Felipe et al. 2014; Henriques et al. 2017). On the other hand, the H-$\beta$ spectra show a broadband intensity enhancement rather than a spectrally localised enhancement, indicating a more distributed perturbation in its formation height.

The absence of distinct signatures in the transition region and solar corona suggests that the associated upflows either dissipate before reaching transition region heights or are too weak to produce measurable brightenings because their energy is radiated off in the chromosphere (Houston et al. 2018; Bose et al. 2019). Similarly, the lack of response in the photospheric layers confirms that flashes are localised to the middle and upper chromosphere, without significant energy deposition in the lower atmosphere.

### 6. CONCLUSION

We investigated flashes in a large solar pore using high-resolution, multi-wavelength observations from SST, IRIS, and AIA. The flashes were spatially confined mostly to the left half of the pore, avoiding the right half with more photospheric structures such as UDs and LB. In the photosphere, flash locations were associated with cooler temperature (ΔT ≈ 400 K), stronger magnetic field (ΔB ≈ 250 G), with greater inclination (∼25° vs ∼18°) compared to the quiescent pore. The flash locations were co-spatial with enhanced power in 3- and 5-minute bands in the photosphere, with only 3-minute power persisting up to the chromospheric layers. The flashes were traced up to 50% line depth layer of the Ca II 8542 Å intensity but not below, with a central upflow (∼1 km s$^{-1}$) of size 1″-2″ and strong downflowing flanks (∼8 km s$^{-1}$). The temperature enhancement associated with flashes was ∼500 K and ∼2500 K at log $\tau$ ∼-5 and -6, respectively. The analysis of 2164 flash spectra showed a bi-modal velocity distribution with ∼52% flashes associated with downflows (∼8 km s$^{-1}$) at log $\tau$ ∼ -5. Having a one-to-one correspondence with flashes, multiple radially outward propagating running waves (5-15 km s$^{-1}$) were observed outside the pore boundary. During propagation, the RWs (amplitude∼1 km s$^{-1}$) steepened and vanished within the first few arcsecs from the pore boundary. Spectral diagnostics showed strong, narrowband emission near the line centre of Ca II 8542 Å and Ca II K lines during flashes, accompanied by apparent polarity reversals in Stokes V for strong flashes. The H-$\beta$ line exhibited a broadband intensity enhancement rather than a narrow core brightening. No distinct signatures of the flashes were detected in photospheric Fe I 6302 Å, IRIS UV or AIA EUV filters, indicating that the flashes are confined to the lower and mid-chromosphere, with no significant impact at the transition region or the coronal heights.

### 7. ACKNOWLEDGMENTS

We acknowledge the community effort devoted to the development of the following open-source Python (python.org) packages that were used in this work: NumPy (Harris et al. 2020), Matplotlib (Hunter 2007), SciPy (Virtanen et al. 2020), and Astropy (Astropy Collaboration et al. 2013, 2018, 2022). This research has used NASA's Astrophysics Data System Bibliographic Services (https://ui.

adsabs.harvard.edu/). RY acknowledges support from NASA ECIP NNH18ZDA001N and NSF CAREER SPVKK1RC2MZ3. TF acknowledges grants PID2021-127487NB-I00, CNS2023-145233, and RYC2020-030307-I funded by MCIN/AEI/10.13039/501100011033.